\documentclass[natbib=false,sigconf,preprint]{acmart}

\copyrightyear{2024}
\acmYear{2024}
\setcopyright{acmlicensed}
\acmConference[ICSE-Companion '24]{2024 IEEE/ACM 46th International Conference on Software Engineering: Companion Proceedings}{April 14--20, 2024}{Lisbon, Portugal}
\acmBooktitle{2024 IEEE/ACM 46th International Conference on Software Engineering: Companion Proceedings (ICSE-Companion '24), April 14--20, 2024, Lisbon, Portugal}
\acmDOI{10.1145/3639478.3640026}

\makeatletter
\def\@copyrightpermission{This is the author's version of the work. It is posted here for your personal use. Not for redistribution. The definitive version was published in ICSE-Companion'24.}
\makeatother

\usepackage{graphicx}
\usepackage{url}
\usepackage{xspace}

\newcommand{\boidae}{\textsc{Boidae}\xspace}

\usepackage{listings}
\lstdefinestyle{default}{
  basicstyle=\ttfamily\footnotesize,
  breaklines=true,
  prebreak=\raisebox{0ex}[0ex][0ex]{\ensuremath{\hookleftarrow}},
  numbers=left,
  numberstyle=\scriptsize,
  xleftmargin=0.4cm
}

\lstdefinestyle{boa}{
	mathescape=false,
	escapechar=\%,
	morestring=[b]`,
	gobble=0,
	morekeywords={_,argument,input,view,exists,foreach,ifall,output,of,weight,stop,visit,before,after,switch,default,current,view},
	emph={int,string,bool,table,float,time,array,stack,map,visitor,%
true,false,%
top,sum,mean,maximum,minimum,set,collection,bottom,%
Project,Person,CodeRepository,Revision,ChangedFile,ASTRoot,Namespace,Declaration,Type,Method,Variable,Statement,Expression,Modifier,IfStatement,%
BooleanLiteral,StringLiteral,PDG,%
InfixExpr,PrefixExpr,PostfixExpr,ForStatement,DoStatement,WhileStatement,TryStatement,ReturnStatement,ThrowStatement,ContinueStatement,BreakStatement,LabeledStatement,%
ExpressionKind,NEW,LITERAL,EQ,NEQ,OP_ADD,OP_MULT,OP_SUB,OP_INC,OP_DEC,BIT_NOT,LOGICAL_NOT,%
TypeKind,CLASS,ANONYMOUS,%
ModifierKind,OTHER,%
ChangeKind,DELETED,%
StatementKind,IF,BREAK,RETURN,THROW,CONTINUE,LABEL,FOR,DO,WHILE,SWITCH,TRY,%
RepositoryKind,SVN},
	emph={[2]logDead,isboollit,isstringlit,isinfix,isprefix,ispostfix,isfixingrevision,getast,iskind,hasfiletype,isliteral,getsnapshot,has_modifier_public,%
new,clear,values,contains,keys,add,format,def,len,match,lowercase,yearof,haskey,remove,strfind,push,pop,peek,normalize,getpdg,getcrypthash,gettotalnodes,gettotaledges,gettotalcontrolnodes,getpdgslice,_row},
}

\lstnewenvironment{Boa}
  {\lstset{language=bash,style=default,style=boa}}
  {}

\usepackage{tikz}
\usetikzlibrary{shapes,arrows.meta,positioning}

\usepackage[inline]{enumitem}
\newlist{enumline}{enumerate*}{1}
\setlist[enumline]{label=(\roman*),itemjoin={{; }},itemjoin*={{; and }}}

\usepackage[style=trad-abbrv,giveninits=true,mincitenames=1,maxcitenames=1,sorting=nty,citestyle=numeric-comp]{biblatex}
\DefineBibliographyStrings{english}{%
  andothers = {\textit{et~al}\adddot}
}
\usepackage[capitalise,nameinlink]{cleveref}

\begin{document}

\title{\boidae: Your Personal Mining Platform}

\author{Brian Sigurdson}
\orcid{0009-0006-5648-6718}
\affiliation{
  \institution{Bowling Green State University}
  \city{Bowling Green}
  \state{Ohio}
  \country{USA}}
\email{bsigurd@bgsu.edu}

\author{Samuel W. Flint}
\orcid{0000-0002-8023-9710}
\affiliation{
  \institution{University of Nebraska-Lincoln}
  \city{Lincoln}
  \state{Nebraska}
  \country{USA}}
\email{swflint@huskers.unl.edu}

\author{Robert Dyer}
\orcid{0000-0001-9571-5567}
\affiliation{
  \institution{University of Nebraska-Lincoln}
  \city{Lincoln}
  \state{Nebraska}
  \country{USA}}
\email{rdyer@unl.edu}



\begin{abstract}
Mining software repositories is a useful technique for researchers and practitioners to see what software developers actually do when developing software. Tools like Boa provide users with the ability to easily mine these open-source software repositories at a very large scale, with datasets containing hundreds of thousands of projects. The trade-off is that users must use the provided infrastructure, query language, runtime, and datasets and this might not fit all analysis needs. In this work, we present Boidae: a family of Boa installations controlled and customized by users. Boidae uses automation tools such as Ansible and Docker to facilitate the deployment of a customized Boa installation. In particular, Boidae allows the creation of custom datasets generated from any set of Git repositories, with helper scripts to aid in finding and cloning repositories from GitHub and SourceForge. In this paper, we briefly describe the architecture of Boidae and how researchers can utilize the infrastructure to generate custom datasets. Boidae's scripts and all infrastructure it builds upon are open-sourced. A video demonstration of Boidae's installation and extension is available at \url{https://go.unl.edu/boidae}.
\end{abstract}

\maketitle

\section{Introduction}
\label{sec:introduction}

Mining software repositories (MSR) is a powerful methodology for discovering developer habits, feature usage, etc. within the Software Engineering (SE) community.  However, performing mining studies comes with great difficulty: generating a dataset takes time, and processing that dataset is difficult.  Libraries like PyDiller~\cite{pydriller} help ease writing mining code but do not provide easy access to the source code's abstract syntax tree (AST) which can either limit your analysis or require additional, complex libraries.  Boa~\cite{Dyer-Nguyen-Rajan-Nguyen-13,boa-website,Dyer-Nguyen-Rajan-Nguyen-15} is a domain-specific query language and runtime infrastructure designed to ease MSR research.  Boa has been shown to reduce the effort necessary to complete MSR studies on ultra-large-scale repositories such as GitHub and SourceForge.  Boa is able to do so by leveraging the Hadoop framework for storage and computation, allowing it to scale to accommodate large datasets.  It provides users with a web-based interface and allows users to easily replicate Boa-based experiments provided by other researchers.

Boa has a few limitations, however.  First, users often want to build customized datasets for their research tasks.  This may be necessary if they want to analyze proprietary datasets or when replicating prior research that requires building a dataset with the same (or comparable) repositories in it, such as was the case for \textcite{keshk23:method-chaining} that built a custom Boa dataset to replicate the prior work of \textcite{nakamaru20}.  Since most researchers do not have access to the Boa infrastructure itself, building a custom Boa dataset is not even an option and they can only query the existing datasets, which currently support languages such as Java, Kotlin, and Python.

The second limitation is in the runtime and query language.  
\textcite{sumon2019pydata} extended Boa to add support for the Python language and produced a dataset of repositories related to data science.  Once again, since most users have no administrative access to the Boa infrastructure, even if they wanted to modify the open-sourced compiler for Boa's query language to add additional features or provide support for other programming languages they want to mine, they currently are unable to do so.

To solve these problems, in this work, we present \boidae.  Whereas Boa is a specific instance of a mining framework, \boidae is a family of (possibly customized) mining frameworks.  While Boa aims to provide an easy-to-use, single place for users wishing to mine open-source repositories, \boidae's intended users are more advanced software miners who need customization not currently feasible with Boa.

Users are able to instantiate their own Boa instance either locally on their computer using Docker\footnote{\url{https://www.docker.com/}} or remotely on a set of cloud servers using Ansible\footnote{\url{https://www.ansible.com/}}.  Thus, if users want to analyze a smaller, but custom, dataset or to test their queries locally on a sample of the larger datasets they can run a Docker instance.  Once they want to scale to a larger dataset, they have the ability to easily spin up a cluster, on any cloud platform they have access to, with their custom Boa dataset.

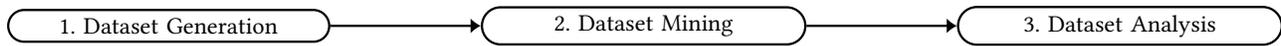
\begin{figure*}[ht]
  \centering
  \begin{tikzpicture}[every path/.style={-{Latex[width=1.5ex,length=1ex]},draw,line width=0.3mm},
    every node/.style={rounded rectangle,draw,align=center},
    node distance=2cm,text width=4cm]

    \node (collection) {1. Dataset Generation};
    \node (mining) [right=of collection] {2. Dataset Mining};
    \node (analysis) [right=of mining] {3. Dataset Analysis};

    \path (collection) -- (mining);
    \path (mining) -- (analysis);
    
  \end{tikzpicture}
  \caption{Overview of the general mining software repositories workflow}
  \label{fig:msr-workflow}
\end{figure*}

The remainder of the paper is organized as follows.
First, we discuss some background on the MSR process and closely related works in \cref{sec:background}.
Next, we discuss \boidae, its use case, workflow, and general use in \cref{sec:boidae}.
We evaluate the approach in \cref{sec:evaluation}.
Finally, we conclude in \cref{sec:conclusion}.

All source code for the Boa language and runtime and \boidae's Ansible and Docker scripts are available on GitHub at \url{https://github.com/unl-pal/Boidae}.

\section{Background and Related Works}
\label{sec:background}

The mining software repositories (MSR) research process can be thought of as having three main stages: dataset generation, dataset mining, and dataset analysis.  The overall workflow is shown in \cref{fig:msr-workflow}.
While all stages can be difficult for new researchers, the first two stages pose several challenges beyond other types of SE research.  First, generating datasets can be time intensive and requires knowledge of things like the GitHub API, how to parse source files with libraries such as Eclipse JDT, and methods for storing large amounts of data such as utilizing distributed filesystems or complex formats such as Protocol Buffers.

Second, even once such a dataset is generated, mining that data also requires substantial expertise.  For example, if the dataset is sufficiently large it might require knowledge of distributed data analysis frameworks such as Hadoop or Spark.  Mining the code structure itself requires a lot of domain expertise and knowledge of things like abstract syntax trees.

Finally, users can analyze the mined data.  Here they may need to leverage things like Pandas for the analysis and Matplotlib for visualizing the results.

All three stages are difficult and time-intensive, however, the first and the second can be abstracted.  Researchers can build and share datasets or tools that enable other researchers to implement their studies more easily.  An example is the Boa infrastructure~\cite{Dyer-Nguyen-Rajan-Nguyen-13,boa-website,Dyer-Nguyen-Rajan-Nguyen-15}, which provides support for the first two steps.  As we will discuss late, \boidae builds on top of Boa.

With Boa, many repositories are collected in advance and translated into a format that is suitable for mining using a parallelizable, domain-specific language.  Boa provides several pre-generated datasets, with support for analyzing several programming languages such as Java, Python, and Kotlin.  The datasets typically have 10's to 100's of thousands of projects each.  Once generated, the datasets are immutable and will never change over time.  This aids the replication of prior results.

Boa also provides a query language designed to abstract away as much of the parallelization as possible.  Users write queries that look sequential, essentially focusing on what data they want to extract from a single project.  Boa then automatically parallelizes the query, running it as a distributed Hadoop program.  This provides scalability, so queries on hundreds of thousands of projects can return in as little as 30 seconds.

There are other prior works that provide some similar functionality.  For example, GHTorrent~\cite{Gousi13,GS12} is a website and dataset containing the event stream from GitHub.  Similar to Boa, users can use a shared infrastructure if they wish to query the dataset, and dataset generation is controlled by the GHTorrent admins.

PyDriller~\cite{pydriller} provides a library to more easily mine Git repositories.  While PyDriller makes it easier to analyze repositories, it does not currently provide support to make the generation of a dataset easier.

Sourcerer~\cite{Linstead:2009} provides a large dataset of Java projects in the form of a SQL database.  Again, the generation of that data is up to the maintainers, and the dataset is provided in a fixed format.  Users must query that data with SQL queries but are free to write any custom mining functions they wish as the infrastructure does not provide a custom query language.

All of these approaches help ease either step 1 or step 2 of the MSR workflow.  But none of them ease both steps while also providing the ability to easily customize the dataset being generated or customize the query language used to mine that data.  This is the main goal of \boidae.

\section{\boidae}
\label{sec:boidae}

Here, we describe \boidae's architecture and an example use case.

\subsection{Use Case}
\label{sec:use-case}

While Boa presents a large dataset and pre-built infrastructure, \boidae allows researchers to build custom datasets and make modifications to the Boa language.  Consider, for example, a researcher who wishes to perform a more one-to-one analysis on a dataset from a prior work.  While they may be able to find and clone all of the same Git datasets as the prior work, writing analyses on it or scaling those analyses could be difficult.

The researcher could write custom code using several different libraries in their programming language of choice, however, this would require keeping track of many small details leading to more complicated code: additionally, it likely would not run quickly as it may not be easily parallelizable.  Using \boidae, our researcher can build a dataset from these projects and run it on any server the research has access to.  Additionally, with some effort, they could extend Boa to include additional information they may be interested in (e.g., output from runs of external tools, or issue tracker data).

\begin{figure}[ht]
  \centering
  \begin{tikzpicture}[every path/.style={-{Latex[width=1.5ex,length=1ex]},draw,line width=0.3mm},
    every node/.style={rounded rectangle,draw,align=center},
    node distance=0.4cm,minimum height=0.8cm,text width=2.8cm]

    \node (user) {\boidae};
    \node (local) [below left=0.5cm and 0.2cm of user] {Docker\\(local/testing)};
    \node (distrib) [below right=0.5cm and 0.2cm of user] {Ansible\\(distributed)};

    \node (local1) [rectangle,below=of local] {Run Docker};
    \node (local2) [rectangle,below=of local1] {\texttt{docker-compose up}};

    \node (dist1) [rectangle,below=of distrib] {Provision Servers};
    \node (dist2) [rectangle,below=of dist1] {Install Ansible};
    \node (dist3) [rectangle,below=of dist2] {Install \boidae};

    \path (user) -- (local);
    \path (user) -- (distrib);
    \path (local) -- (local1);
    \path (local1) -- (local2);
    \path (distrib) -- (dist1);
    \path (dist1) -- (dist2);
    \path (dist2) -- (dist3);

  \end{tikzpicture}
  \caption{Overview of the \boidae architecture}
  \label{fig:arch}
\end{figure}
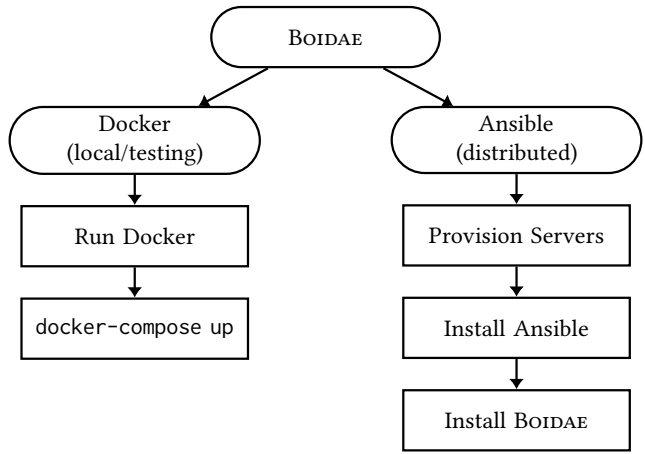

\subsection{Architecture}
\label{sec:architecture}

The architecture of \boidae is split, based on two different runtime environments, as shown in \cref{fig:arch}.  First, we provide a Docker image for running \boidae locally on a single machine.  This image is designed to support two use cases: testing and analyzing small datasets.  Users utilize Docker's compose functionality to run the container and then connect to a web service running on \texttt{localhost}.  This provides a web interface similar to the one Boa provides, allowing the user to select a dataset and run a query.

The container also contains helper scripts for generating customized datasets.  The scripts provide support for both GitHub and SourceForge, or the user can simply manually clone any Git repositories and generate a dataset from that set of repositories.  If generating from GitHub-based projects, the user has the option of simply specifying a list of repositories they want to generate a dataset from or they can use more generic tools that will use GitHub's API to search for projects matching the user's requirements.

The container also provides support for modifying the runtime.  The language's compiler and runtime infrastructure are installed in source format, with a full build environment available.  There are also support scripts to help update/install new versions of the compiler, once users have modified it.

The second supported runtime environment is a distributed environment.  This helps support any use case where the user has a larger dataset they might need to query frequently.  For this, the user needs to first provision a set of compute nodes.  This can be done with their employer's/university's computing services, with a cloud such as Amazon Web Services (AWS)~\cite{AWS}, or using research infrastructure such as CloudLab~\cite{CloudLab} or ACCESS~\cite{access}.  Once they have servers provisioned, they can install Ansible and use our provided Ansible scripts to get a \boidae distributed instance installed and configured.  They still have the ability to customize the dataset or runtime when utilizing the Ansible scripts.

\subsection{Workflow}
\label{sec:workflow}

With \boidae, the overall workflow remains the same as shown in \cref{fig:msr-workflow}, but there are changes in some details, and steps 2 and 3 are more easily repeated.
Continuing on with our prior use case, our researcher would complete the following steps.

First, a basic Boa cluster would be configured using either the Ansible scripts or the Docker file.
The Ansible scripts will allow for the installation of a \boidae cluster on a number of nodes, though it is specifically designed for CloudLab \cite{CloudLab}.
The Docker file is instead useful for the generation of datasets, testing of modifications to the compiler on small datasets, or general work on small datasets.
At this point, any modifications to the dataset generation code (to store results of external tools), the compiler (for further integration of external tools), or the run time (addition of language-level functions) should be completed.

Then, they would collect metadata files for each repository.
These are JSON files that are collected using the GitHub API, which describe a number of properties of the project, including star gazer count.
These JSON files are then used to generate the dataset through the use of one of several scripts.
This script will clone the repositories automatically.
To ensure that private repositories can be cloned appropriately, GitHub's HTTPS authentication must be correctly configured.
Generation of a dataset will take time when many repositories are ingested.
Once the generation is complete, the dataset can be installed into the configured \boidae cluster.
From this point forward, steps 2 and 3 may commence: the dataset is now ready to be used for studies.

Now the dataset may be mined using the Boa language, running queries like that shown in \cref{fig:boa-query}, which counts the number of annotations per project.
Writing Boa queries is a topic in itself, but generally, the visitor pattern is used to visit each commit, and within each commit, each changed file.
A query can drill down to whatever level of the file is necessary, and tabular data is generated.
The example query shows data indexed by project, but indexing by other elements is possible.

\begin{figure}[htbp]
  \centering
\begin{Boa}
o: output sum[project: string] of int;

visit(input, visitor {
    before node: CodeRepository -> {
        snapshot := getsnapshot(node);
        foreach (i: int; def(snapshot[i]))
            visit(snapshot[i]);
        stop;
    }
    before mod: Modifier -> {
        if (mod.kind == ModifierKind.ANNOTATION)
            o[input.id] << 1;
    }
});
\end{Boa}
  \caption{Count number of annotations per project.}
  \label{fig:boa-query}
\end{figure}

These tabular data files can be easily converted to CSV and analyzed using any preferred tools.
Support for building multiple-file analyses is partially provided by the Boa Study Template.\footnote{\url{https://github.com/boalang/study-template}}

\section{Evaluation}
\label{sec:evaluation}

We evaluated the \boidae infrastructure in four ways.  First, we wanted to ensure that the infrastructure can run on multiple systems.  To evaluate this, we first verified we can run the Docker version of the infrastructure on several different machines.  Next, we wanted to verify if the cloud version of \boidae works.  To verify this, we used the Ansible scripts to install \boidae onto a small cluster of five nodes at Bowling Green State University.  We also utilized the CloudLab~\cite{CloudLab} infrastructure, an experimental testbed that allows researchers to experiment with cloud infrastructures.  Once we configured the server infrastructure, we were able to once again utilize the Ansible scripts to get an instance of \boidae running on CloudLab's servers.

Our second evaluation was to see if it is possible to build custom datasets.  We used \boidae's feature for building GitHub datasets to build a custom dataset containing the repositories in the \texttt{boalang} user, which represents the open-source infrastructure for Boa.  We were able to successfully generate the dataset and run sample queries on it.

The third evaluation is to see if \boidae supports customizing the language and runtime.  For this evaluation, we modified the Boa language to add additional support for custom domain-specific mining functions.  We then verified that those custom functions were available in queries and manually inspected the results to verify they worked as intended.

\begin{figure}[ht]
    \centering
    \includegraphics[width=\linewidth]{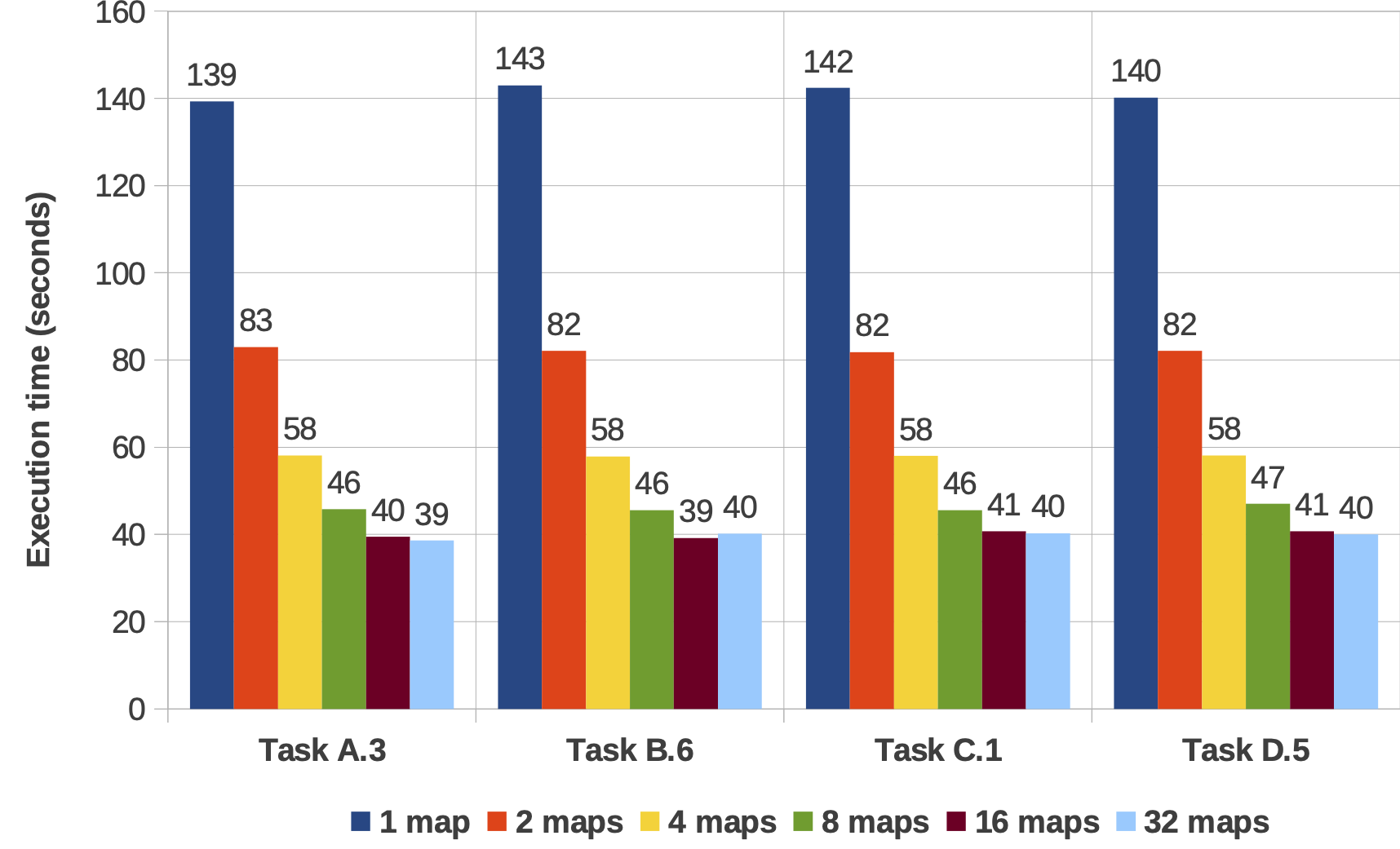}
    \caption{Task execution times as the number of maps increases \cite{Dyer-Nguyen-Rajan-Nguyen-13}.}
    \label{fig:scalability}
\end{figure}

Finally, we previously evaluated the scalability of the infrastructure.  Since the infrastructure runs on top of Hadoop, we ran several tasks in a prior work~\cite{Dyer-Nguyen-Rajan-Nguyen-13}.  We then varied the number of map tasks from 1 to 32.  See Figure~\ref*{fig:scalability} for the results.

As can be seen from the figure, the time needed to execute the queries decreases as the number of parallel map tasks increases.  It then levels out at around 16 map tasks.  This shows that the infrastructure can scale, providing good performance.

\section{Conclusion}
\label{sec:conclusion}

Mining software repositories is an important field of research in SE.  Tools like Boa provide ways to more easily mine repositories but, despite being open-sourced, do not allow users the ability to customize the dataset or the runtime.  \boidae provides a solution to allow researchers to run their own customized instances of Boa, enabling them to easily generate and mine their own custom datasets.  \boidae provides the ability to run locally on Docker, or scale to cloud-scale using Ansible scripts.

An operational Boa instance is available at \url{https://boa.cs.iastate.edu/boa/} and available to the public.
For anyone wishing to customize either the dataset or Boa's infrastructure, \boidae provides installation in the form of either Ansible scripts at \url{https://github.com/boalang/ansible} or a Docker container at \url{https://github.com/boalang/boa-docker}.  The Boa infrastructure, which \boidae utilizes, is available at \url{https://github.com/boalang/compiler} and \url{https://github.com/boalang/drupal}.

All artifacts for this demo are available in GitHub at \url{https://github.com/unl-pal/Boidae}.  A video demonstration of Boidae's installation and extension is available at \url{https://go.unl.edu/boidae}.

\begin{acks}
  This work was supported in part by the \grantsponsor{NSF}{U.S. National Science Foundation (NSF)}{https://www.nsf.gov/} under grants \grantnum{NSF}{1512947} and \grantnum{NSF}{1518776}.
\end{acks}

\printbibliography

\end{document}